\documentclass[aps,twocolumn,prb,showpacs]{revtex4-1}
\usepackage{graphicx}
\usepackage{color}
\usepackage{amsmath}
\usepackage[bf,Large]{}
\usepackage{subfloat}

\usepackage{natbib}
\usepackage{longtable}
\usepackage{dcolumn}
\usepackage{tabularx}
\usepackage{keyval}
\usepackage{multirow}

\begin{document}

\title{Effect of orbital symmetry of the tip on Scanning Tunneling
  Spectra of Bi$_2$Sr$_2$CaCu$_2$O$_{8+\delta}$}

\author{Ilpo Suominen}
\affiliation{Department of
Physics, Tampere University of Technology, P.O. Box 692, FIN-33101
Tampere, Finland}

\author{Jouko Nieminen}
\email{jouko.nieminen@tut.fi}
\affiliation{Department of
Physics, Tampere University of Technology, P.O. Box 692, FIN-33101
Tampere, Finland~}
\affiliation{Department of Physics, Northeastern
University, Boston}

\author{R.S. Markiewicz} \author{A. Bansil}
\affiliation{Department of Physics,
Northeastern University, Boston}

\date{Version of \today}

\begin{abstract} We discuss how variations in the scanning tunneling
  microscope (STM) tip, whether unintentional or intentional, can lead
  to changes in topographic images and dI/dV spectra. We consider the
  possibility of utilizing functionalized tips in order to improve the
  sensitivity of STM experiments to local irregularities at the
  surface or hidden below the surface layers. The change in the tip
  symmetry can radically alter the contrast of the topographic image
  due to changes in tip-surface overlap. The dI/dV curves change their
  shape according to which sample bands the tip orbital tends to
  overlap. In addition, relative phases between competing tunneling
  channels can be inverted by changing the tip symmetry, which could
  help reveal the origin of a local irregularity in tunneling spectrum.
\end{abstract}

\date{Version of \today}

\maketitle

\section{Introduction}

Scanning tunneling microscopy and spectroscopy (STM/STS) are
extensively used to probe the quasiparticle spectra of high
temperature superconductors, especially of the cuprate materials such as
Bi$_2$Sr$_2$CaCu$_2$O$_{8+\delta}$
(Bi2212)\cite{Fischer,McElroy,Hudson,Pan, Yazdani}. These studies have
contributed immensely in obtaining insight into superconducting and 
pseudogap phases
of these materials owing to the very high spatial and energy 
resolution of the spectroscopy.  The full experimental potential of
STM/STS is, however, far from exploited, since the chemical and
spatial resolution as well as the directional selectivity could, in 
principle, be
improved by overcoming various challenges related to intrinsic
properties of STM.

The first challenge is related to the fact that the signal from
superconducting layers is filtered by insulating oxide layers. In
addition, the signal does not arrive at the microscope tip directly
from the site below the tip but rather through symmetry driven
tunneling channels. Hence, experimentally observed spectral features
are not only distorted by filtering, but ascertaining their spatial
origin becomes harder the deeper within the sample is their
source. Another factor to be aware of is that the electronic and
geometrical properties of the microscope tip may also affect STM
images. In standard interpretation, the tip is assumed symmetric to
rotations around the vertical axis (`s-wave'). The relative bluntness
of the tip suggests this assumption and, furthermore, the metallic
nature of common tip materials does not favor oriented ``dangling''
orbitals, and specific effects of tip geometry and electronic
structure are usually ignored.  However, this interpretation has been
challenged by a variety of sophisticated theoretical studies
\cite{cjchen,chaika,sacks}, which, on the one hand show that the
closer the tip is to the surface, the more important the symmetry of
tip orbitals is to the STM image and, on the other hand, they indicate
that s-wave tips might not be able to produce images with atomic
resolution. Notably, the relative importance of tip orbitals with
different symmetries in the STM signal can be controlled by varying
the tip-surface distance, or alternatively, by functionalizing the tip
by attaching particular atoms/molecules at the apex of the tip.

In various experimental
studies\cite{eigler,hahn,bartels,bartels_prl_1998,lee_science_1999,
lagoute_prb_2004,repp_prl_2005,deng,cheng}, the tip properties have
been deliberately modified for better spatial or chemical selectivity
by picking up a molecule, e.g. CO, O$_2$ or N$_2$, from the surface to
the apex of the tip after which the plain tip results have been
compared to the results with a functionalized tip.  This has proven to
improve chemical selectivity in adsorbate on a solid surface system
since each surface species responds in an individual way to the change
of tip symmetry due to functionalization. Even in the case of
relatively complicated molecules, such as pentacene attached to the
tip, the contrast of differential conductance maps revealing the
geometric shape of molecular orbitals is surprisingly sharpened
\cite{repp_prl_2005} by the enhanced directionality of the probe-sample
overlap due to the dominance of the $\sigma$ or $\pi$ type localized
orbitals closest to the sample surface.  The idea of utilizing a
functionalized tip is further encouraged by the fact that the
resolution of atomic force microscopy (AFM) studies is enhanced by
using CO molecule attached to the metallic tip \cite{AFM_CO}, although
improving the resolution and selectivity in the case of AFM has a
different physical basis from STM.

The preceding work on molecular adsorbate systems provides motivation
for considering functionalized tips for investigating cuprate
materials, where the interesting physics is found at layers not
immediately on the surface. We can foresee many uses of functionalized
tips in probing these materials. An obvious application would be a
more accurate probe for the surface properties. In the case of Bi2212,
to date one has mostly observed the Bi atoms of the surface layer in
topographic STM images. Information from the oxygen atoms of the
surface layer has been difficult to obtain, and hence STM provides
limited ability to observe the real symmetry of the surface. A more
involved application is related to the dI/dV spectra. As shown in our
previous studies\cite{nieminenPRB}, the spectra are mappings of the
electronic structure of the cuprate layer, filtered by the BiO and SrO
layers, which sit on top of the cuprate layers. Although the signal
can be theoretically resolved into partial signals of different
origin, the STS experiment measures a signal combined from different
sources and, furthermore, the information from different directions in
momentum space is integrated into a single outgoing signal. A possible
means of decomposing the STS spectrum experimentally is to measure
topographic spectral maps. In practice, real space and Fourier
transformed maps differentiating electron and hole states have been
produced in order to recognize direction dependent underlying
electronic order \cite{z_etc_maps}. A third possible application would
be investigation of symmetry properties related to nematicity
transitions in the subsurface electronic structure at low values of
doping, where the oxygen atoms of CuO$_{2}$ layers in different
directions appear non-equivalent below the pseudogap
transition temperature $T^{*}$ as shown in Ref. \onlinecite{Kohsaka2007}.\\

In fact, functionalized tips may well have been used inadvertently in
studies of Bi2212. It has been reported that tip resolution can
improve dramatically after use, and it has been suggested that this
may be due to the attachment of a Bi atom to the tunneling tip
\cite{bifunct}.  Here we will explore the effects of tips, which are
dominated by a single atomic orbital of varying symmetry.  We find
that such tips can not only couple with different surface orbitals,
but can also affect the interference between tunneling channels from
the cuprate layer to the microscope tip. This suggests possibilities
for investigating irregularities within the cuprate layers, which are
not directly accessible to the STM tip, and thus reveal features
invisible to STM/STS experiments with a symmetric tip. In this way,
our study advances the understanding of matrix element effects in
STM.\cite{abfoot1,newarpes,rixs,Compton,positron}

This article is organized as follows. Section II briefly describes the
model utilized in this study. Section III presents calculations of the
effect of tip symmetry on topographical maps, STM corrugations and
dI/dV spectra at various high symmetry sites. In Section IV we show
how a hidden irregularity within the cuprate layer can be studied by
comparing results obtained with different tip symmetries, among other
results.  Finally, conclusions are given in Section V.

\section{Description of the model}

\begin{figure}
  \includegraphics[width=0.5\textwidth]{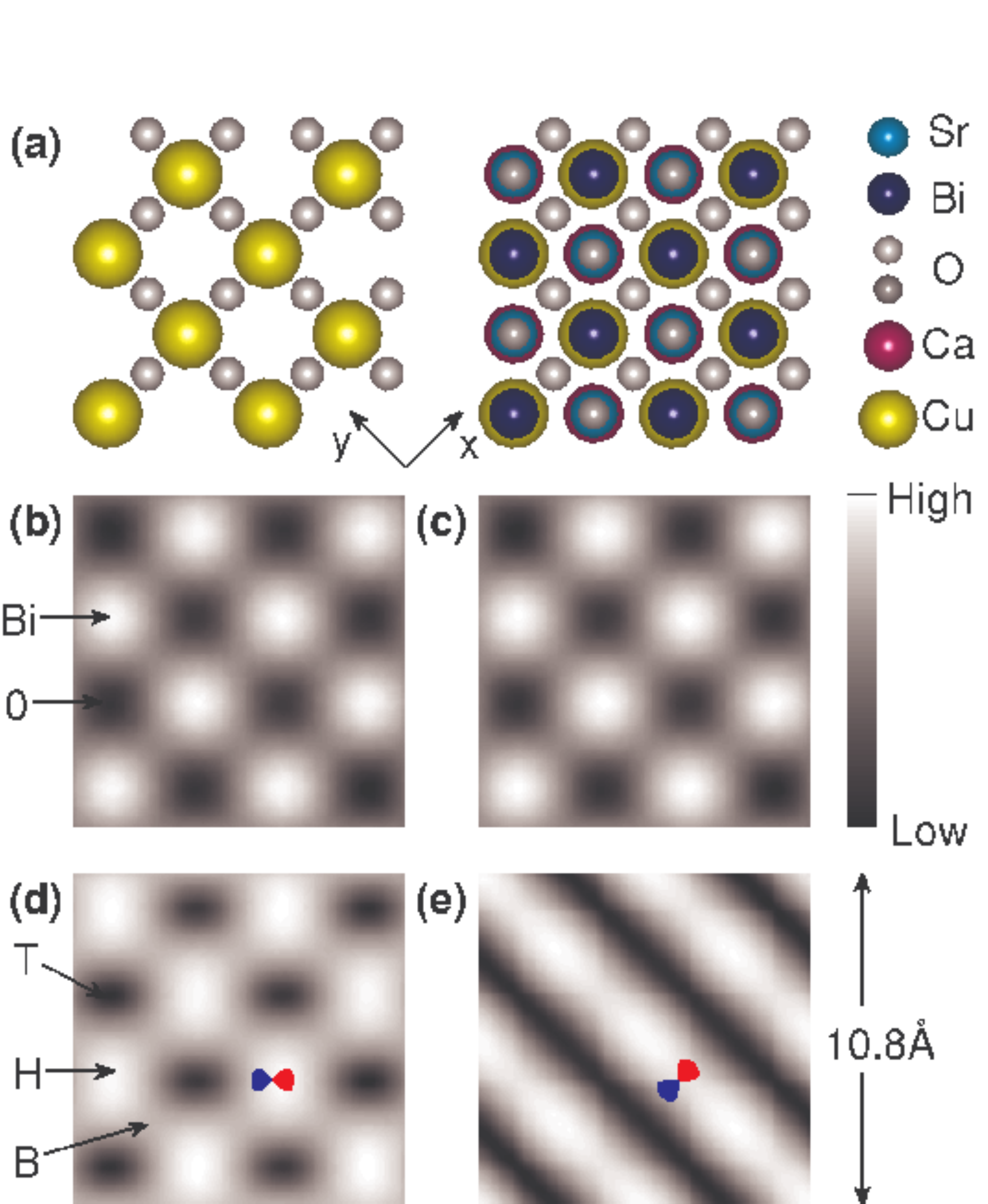}
  \caption{ (color online) (a) Geometry of the simulated system. Left
    hand panel gives the arrangement of the CuO$_{2}$ layer, while the
    right panel depicts the BiO and SrO layers on top of the CuO$_{2}$
    layer; (b)-(e) Topographic STM images for different tip
    symmetries. (b) $s$-orbital, (c) $p_{z}$-orbital, (d)
    $p_{r}$-orbital, (e) $p_{b}$-orbital.  Indicated also are Bi and O
    positions and the top (T), bridge (B) and hollow (H) sites. In (d)
    and (e) the tip symmetry is further indicated with a schematic of the
    corresponding p-orbital.}
  \label{topo}
\end{figure}

Although the model used is essentially the same as in
Ref. \onlinecite{nieminenPRB}, we provide an overview here for the
sake of completeness. The simulation supercell consists of 8 unit
cells that have been rotated $45$ degrees as shown in
Fig. \ref{topo}. The unit cell geometry is adapted from
Ref. \onlinecite{Bellini}. In total there are $120$ atoms and
$2\times464$ orbitals accounting for both spins.  The topmost 7 layers
below the cleavage plane are included, which are in descending order:
$BiO$, $SrO$, $CuO_{2}$, $Ca$, $CuO_{2}$, $SrO$ and $BiO$.  Depending
on the atom type different sets of orbitals are included. For $Bi$,
$O$ and $Ca$ the orbital set is $(s,p_x,p_y,p_z)$ and for $Cu$ the
orbital set is
$(4s,d_{3z^{2}-r^{2}},d_{xy},d_{xz},d_{yz},d_{x^{2}-y^{2}})$. $Sr$ has
a single $s$ orbital. The Green's function is computed in a supercell
Brillouin zone at $256$ equally distributed {\bf k}-points
corresponding to $8\times256=2048$ {\bf k}-points in the unit cell.

As shown in Fig. \ref{topo}, we define the position of Cu (or
surface Bi) atoms as top site ($T$), the position of bonding O atoms
of the cuprate layer as bridge site ($B$), and the position of O(Bi)
as the hollow site ($H$). In this choice of geometry, the $p_{x}$ and
$p_{y}$ orbitals point along the bond between the neighbouring Cu
atoms, i.e., in the direction from the top site to the bridge site ($T
\rightarrow B$), whereas the linear combinations $\frac{1}{\sqrt{2}}
(p_{x} \pm p_{y})$ point from the top site towards the hollow site ($T
\rightarrow H$). Concerning the oxygen atoms related to the cuprate
layer, the {\it apical oxygen} is at the top site, while the {\it
bonding oxygens} between Cu atoms are at bridge sites.

The tip is modeled as a single orbital in order to
highlight the effect of each type of tip symmetry on the STM/STS
spectrum.  We consider four different tip symmetries which can be
divided into two groups. These consist of $s$- and $p_{z}$-orbitals,
which are symmetric to rotation around z-axis, accompanied with two
choices of horizontal $p$-orbitals.  We use a shorthand notation
$p_{b}$ for a $p$-orbital along the bond between two neighbouring Cu
atoms, and $p_{r}$ for an orbital rotated by $45^{\circ}$, i.e.,
aligned in the direction connecting next nearest neighbouring Cu
atoms. These two horizontal tip symmetries can also be connected to the
nodal ($p_{r}$) and antinodal ($p_{b}$) directions.

The Hamiltonian is the same as that used previously in our STM studies
of optimally doped B2212\cite{NLMB,nieminenPRB,selffoot,tanmoyop},
where superconductivity is modeled with a pairing matrix
$\Delta_{\alpha \beta}$ between electrons and holes with opposite
spins:
\begin{equation}
  \begin{array}{ccc}
\hat{H}& = & \sum_{\alpha\beta\sigma}
\left[\varepsilon_{\alpha}c^{\dagger}_{\alpha \sigma} c_{\alpha \sigma}+
V_{\alpha \beta}
c^{\dagger}_{\alpha \sigma} c_{\beta\sigma}\right]\\
&+&
\sum_{\alpha \beta
\sigma} \left[\Delta_{\alpha \beta} c^{\dagger}_{\alpha \sigma}
c^{\dagger}_{\beta -\sigma} + \Delta_{\beta \alpha}^{\dagger}
 c_{\beta -\sigma} c_{\alpha \sigma} \right]
  \end{array}
\label{hamiltonian}
\end{equation}
with real-space creation (annihilation operators) $c^{\dagger}_{\alpha
  \sigma}$ ($c_{\alpha \sigma}$).  Indices are combination of site and
spin indices.  The d-wave symmetry of the superconducting gap is
modeled by taking $\Delta$ to be non-zero only between the
neighbouring $d_{x^{2}-y^{2}}$ orbitals, and with opposite signs for
$x$- and $y$-directions, which leads to a d-wave symmetric
superconducting gap \cite{Flatte}. The tight-binding parameters
$V_{\alpha \beta}$ are taken to be doping-independent in the spirit of
the rigid band model.\cite{abfoot2,new1,new2}

The tunneling current is computed within the Todorov-Pendry
approximation
\cite{Todorov,Pendry,tunnelingfoot,meirwingreen,TH,korventausta} as
\begin{equation}
\sigma = \frac{dI}{dV} = \frac{2 \pi e^{2} }{ \hbar} \sum_{t t' s s'}
\rho_{tt'}(E_F)V_{t's} \rho_{ss'}^{}(E_F+eV)V_{s't}^{\dagger}.
\label{conductance}
\end{equation}
   
The hopping integrals $V_{\alpha\beta}$ depend on the symmetry of tip
and surface orbitals as well as the distances involved through the
Slater-Koster coefficients\cite{Slater, Harrison}. The tip to surface
distance is usually greater than the distance between the atoms in the
bulk material. To allow for this the standard Slater-Koster method is
enhanced by an exponential decay at large distances
\cite{hoppingfoot}. The Slater-Koster method allows one to take the
directionality of both the tip and surface orbitals into account. It
can be considered as a special case of the more general ``derivative
rule'' of Chen\cite{cjchen} and a similar approach of Sacks and
Noguera \cite{sacks} who utilize derivatives of the real space tip
wave function.

There is an alternative way of writing Eq. \eqref{conductance}:
\begin{equation}
\sigma = \frac{2 \pi e^2 }{ \hbar} \sum_{t t' c c'}
\rho_{tt'}(E_F)M_{t'c} \rho_{cc'}^{}(E_F+eV)M_{c't}^{\dagger},
\label{mconductance}
\end{equation}
where
\begin{equation}
  M_{tc} = V_{ts}G^{0+}_{sf}V_{fc}.
  \label{filter}
\end{equation}

The difference between the two formulae is that
Eq. \eqref{conductance} emphasizes the coupling between the tip and
the surface orbitals expressed by the hopping integrals $V_{ts}$,
whereas Eq. \eqref{mconductance} makes more transparent how tunneling
takes place through the filtering layers between the tip and the
superconducting layers.  Of course the symmetry of the tip itself is
described by $V_{ts}$, but in order to recognize the relevant surface
orbitals, it is essential to consider the filtering matrix elements,
$M_{tc}$, as well.

The spectral function $\rho_{cc'}$ depends on both regular and anomalous 
Green functions as shown below
\begin{equation}
  \rho_{cc'}=-\frac{1}{\pi}\sum_{\alpha}(G^{+}_{c\alpha}\Sigma{''}_{\alpha}G^{-}_{\alpha
    c'}+F^{+}_{c\alpha}\Sigma{''}_{\alpha}F^{-}_{\alpha c'})
  \label{dmatrix}
\end{equation}
To conclude, the formulae given above provide not only the total $dI/dV$
spectrum, but also an explicit form of the filtering effect on the
cuprate layer, and decompose the density
matrix of the cuprate orbitals into regular and anomalous spectral
terms.

\section{Results}

\subsection{Topographic maps}

\begin{figure}
  \includegraphics[width=0.4\textwidth]{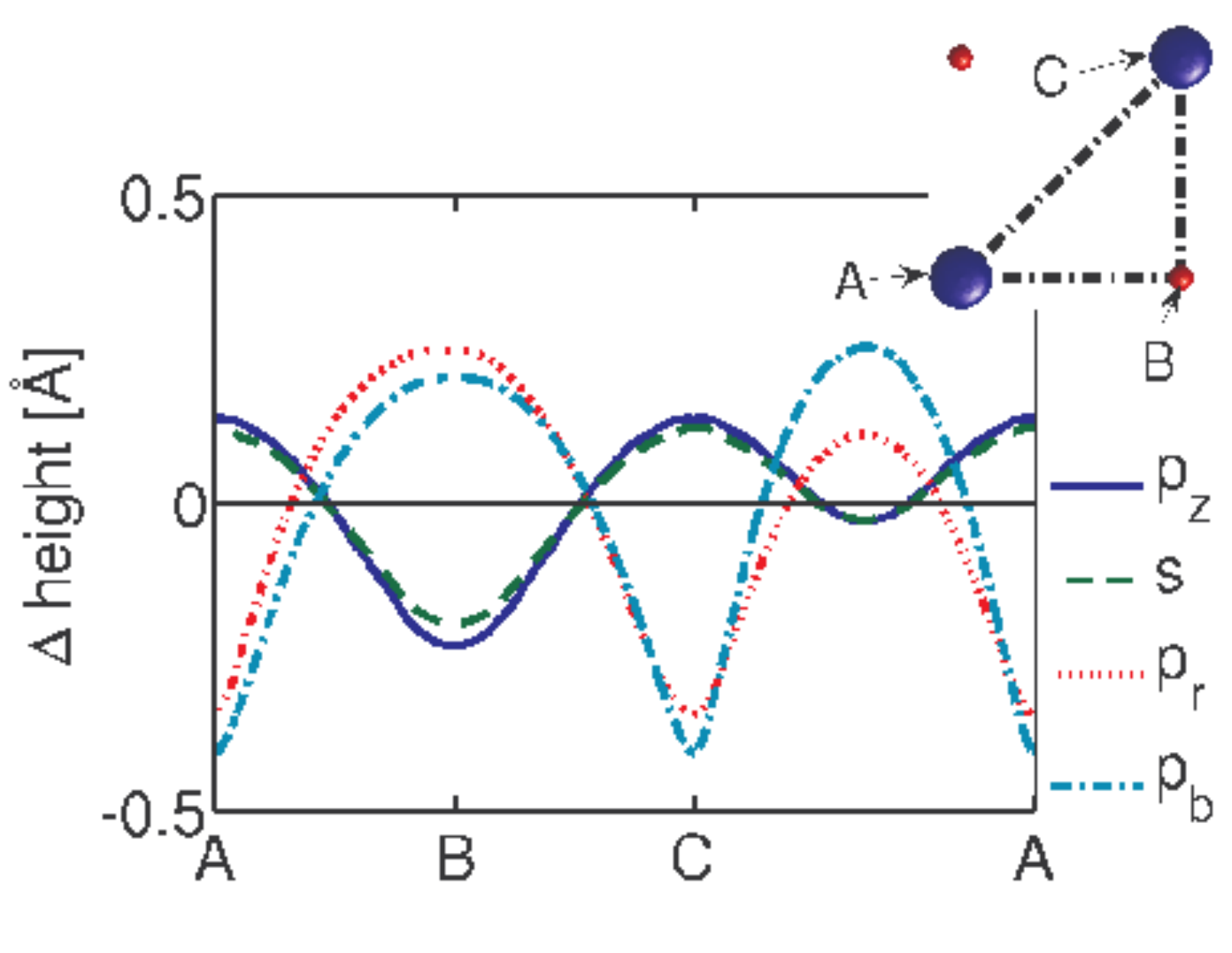}
  \caption{(color online) Corrugation curves for various tip
    orbitals along the route ($A \rightarrow B \rightarrow C
    \rightarrow A$) shown in the insert. The vertical scale represents
    the height change the tip experiences along the route. 
Curves for $p_{r}$ and $p_{b}$ are similar in
    overall shape yet they differ strongly in relative amplitude. This
    shows that they connect at least partially to different surface
    orbitals. Note that each curve was shifted downwards by its own average 
    tip surface distance thus making all curves zero centered.}
  \label{corrugation}
\end{figure}

The topographic maps of Fig. \ref{topo}(b-e) are calculated using
Eq. \eqref{conductance} in constant current mode, i.e., there is a
feedback loop varying the tip-surface distance in order to retain the
tunneling current constant. For bias voltage we choose $-0.3 V.$ We
also follow the common practice of accompanying topographic maps
with corrugation curves in Fig. \ref{corrugation}, since the latter
reveal the absolute scale of the contrast otherwise obtained in a
rather arbitrary way from topographic maps.

Experimental topographic maps typically resemble the calculated maps
for s-wave tip with a high intensity at the position of the surface Bi
atoms.  Within the Tersoff-Hamann type analysis \cite{TH}, a high
intensity signal is attributed to a high local density of
electrons. As Fig. \ref{LDOS}(a) indicates, the surface Bi atoms have
a large density of states over a wide range of energies around the
Fermi energy. This was demonstrated already in our earlier work
\cite{nieminenPRB} where we further related this to the coupling
between the $p_{z}$ orbitals of Bi and the $d_{z^{2}}$ orbital of the
Cu atom below through the intermediate $p_{z}$ orbital of the apical
oxygen. In contrast, orbitals of O(Bi) contribute to the BiO bands,
which contain a gap around the Fermi-energy.  In Fig. \ref{LDOS}(a)
these states are broadened since we have invoked a Fermi liquid type
self-energy in the noncuprate layers \cite{ldafoot}.

As seen in Figs. \ref{topo} (b) and \ref{topo} (c), tips with $s$- as
well as $p_{z}$-symmetry expose the Bi atoms as bright spots. This is
hardly surprising, especially in the case of the rather strongly
oriented $p_{z}$ tip orbital. Our computations also suggest that
$p_{z}$ symmetry slightly sharpens the contrast of the topographic
map, but as shown in the corrugation curves of Fig. \ref{corrugation},
the effect may not be experimentally significant. Tips with in-plane
p-symmetry show striking changes in topographical maps. In the case
where the tip orbital lies along the top-hollow (nodal) direction,
$p_{r}$ (Fig.  \ref{topo}(c)), the contrast is inverted so that O(Bi)
atoms appear as bright spots. However, as discussed further in Section
IV below, visibility of oxygen atoms is spurious since the tunneling
signal is still dominated by overlap between the horizontal tip
orbital and the $p_{z}$ orbitals of Bi atoms. This is also the obvious
reason why the bright spots in the topographical image lose their
sharpness.  The change in contrast becomes even more dramatic in the
case of the tip orbital lying along the top-bridge (anti-nodal)
direction, $p_{b}$. As seen in Fig. \ref{topo} (d), the fourfold
symmetry is now broken, and bright lines appear perpendicular to the
orbital direction. In addition, intensity maxima are seen at bridge
sites when the bonding oxygen orbital points to the same direction as
the tip orbital. This indicates that in this case also the nature of
the topographic image is driven the overlap between the tip orbital
and the $p_{z}$ orbital of Bi atoms.

The intensity patterns in topographic images can be quantified as
corrugation curves, such as those shown in Fig. \ref{corrugation}.
Interestingly, the horizontal tip orbitals not only invert the
contrast of the topographic image, but they are also more
sensitive to the horizontal position than tips with rotational
symmetry around the z-axis. This is seen in the more strongly varying
corrugation as the tip is scanned between the top, bridge and hollow
sites. These differences, originating from the symmetry dependent
nature of the tip-surface overlap, might be useful in pursuing a
better lateral resolution in STM imaging.

Unfortunately, our theoretical predictions cannot be compared directly
with measurements at this time since we are not aware of any
experimental STM literature on Bi2212 where a deliberately
functionalized tip has been deployed. However, some insight can be
obtained by considering a study such as that of
Ref. \onlinecite{cheng}, which discusses the effects of attaching
O$_{2}$ functionalizing molecules to a tungsten tip for investigating
a metal phtalocyanine monolayer on an Au(111)
surface. Ref. \onlinecite{cheng} finds selectivity with respect to
molecular orbitals between normal and O$_{2}$-functionalized tips, and
observes the STM pattern of the adsorbate monolayer to rotate as
different tips are used to probe the surface.

\subsection{dI/dV spectra}

\begin{figure}[th]
 
  \begin{minipage}{4.25cm}%
    \includegraphics[width=1.0\textwidth]{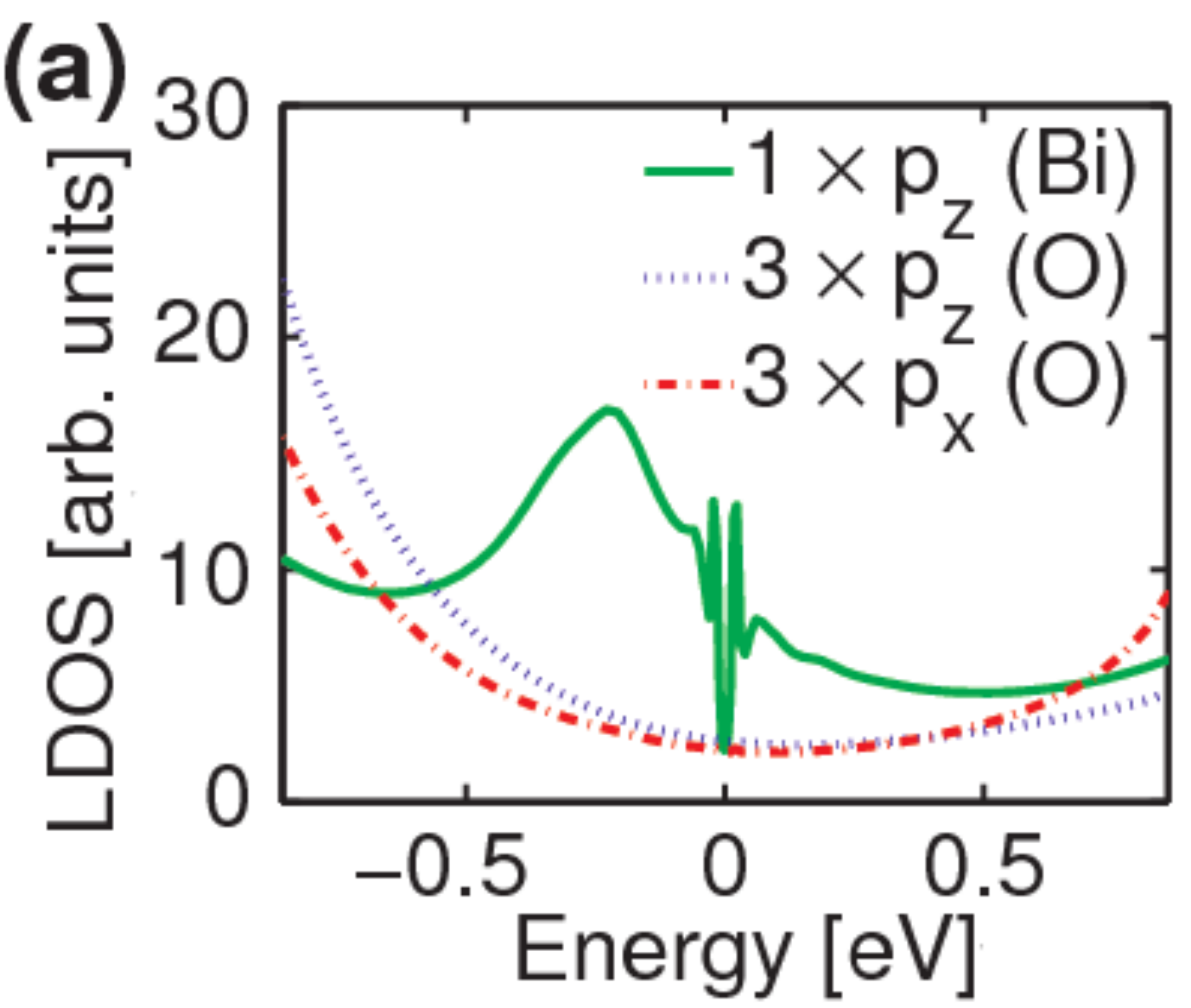}
  \end{minipage}%
  \begin{minipage}{4.25cm}%
    \includegraphics[width=1.0\textwidth]{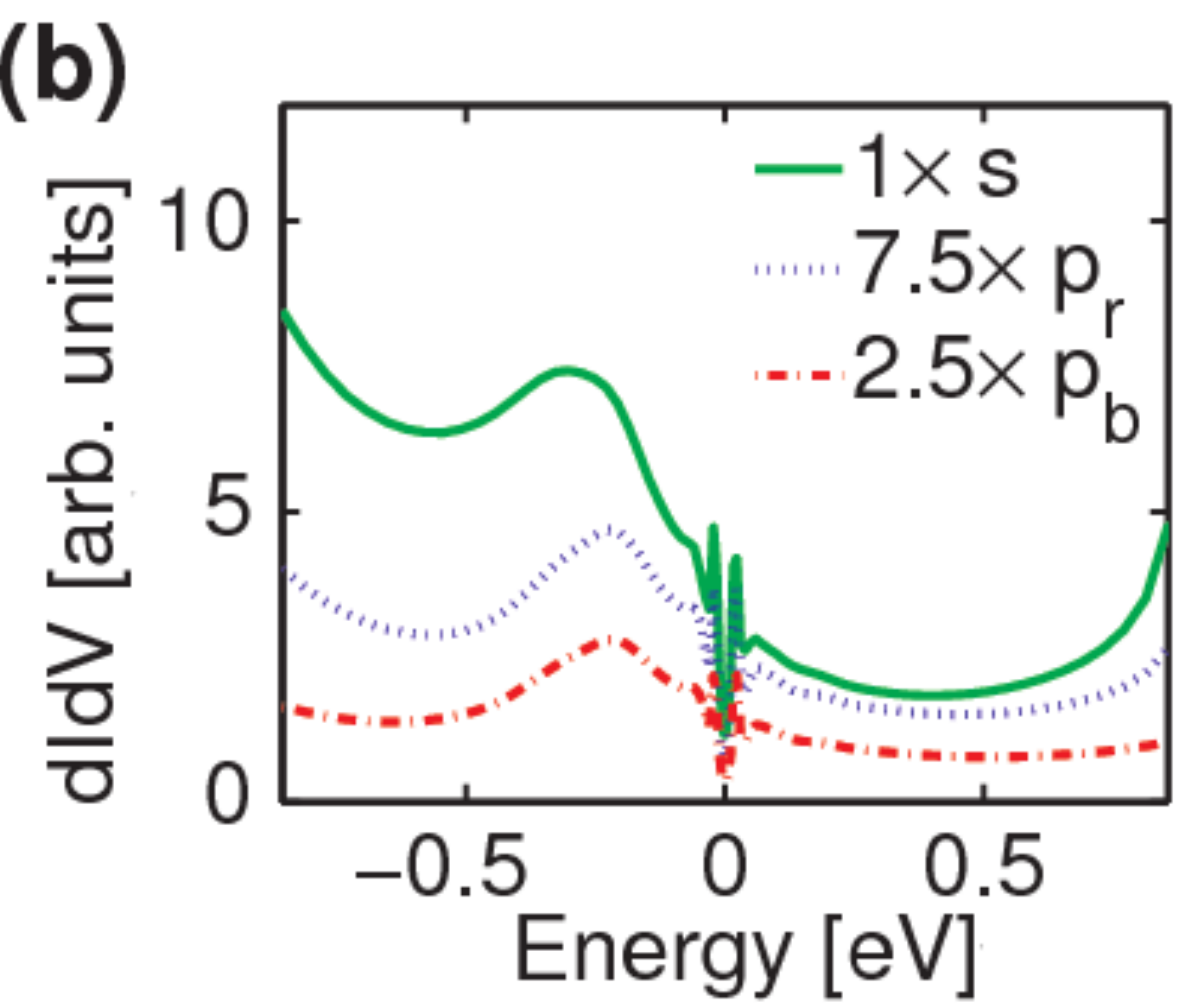}
  \end{minipage}\\
  \begin{minipage}{4.25cm}%
    \includegraphics[width=1.0\textwidth]{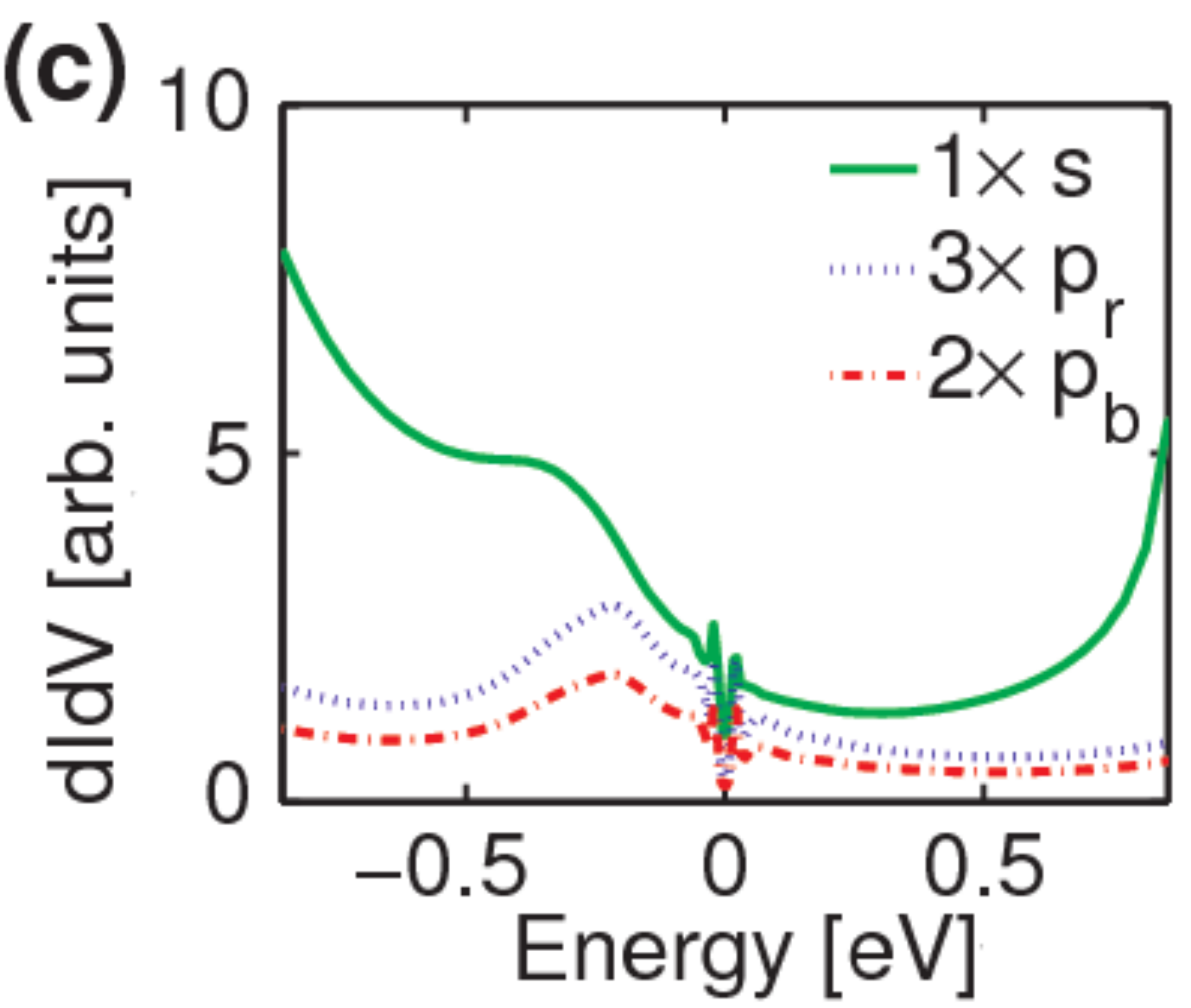}
  \end{minipage}
  \begin{minipage}{4.25cm}%
    \includegraphics[width=1.0\textwidth]{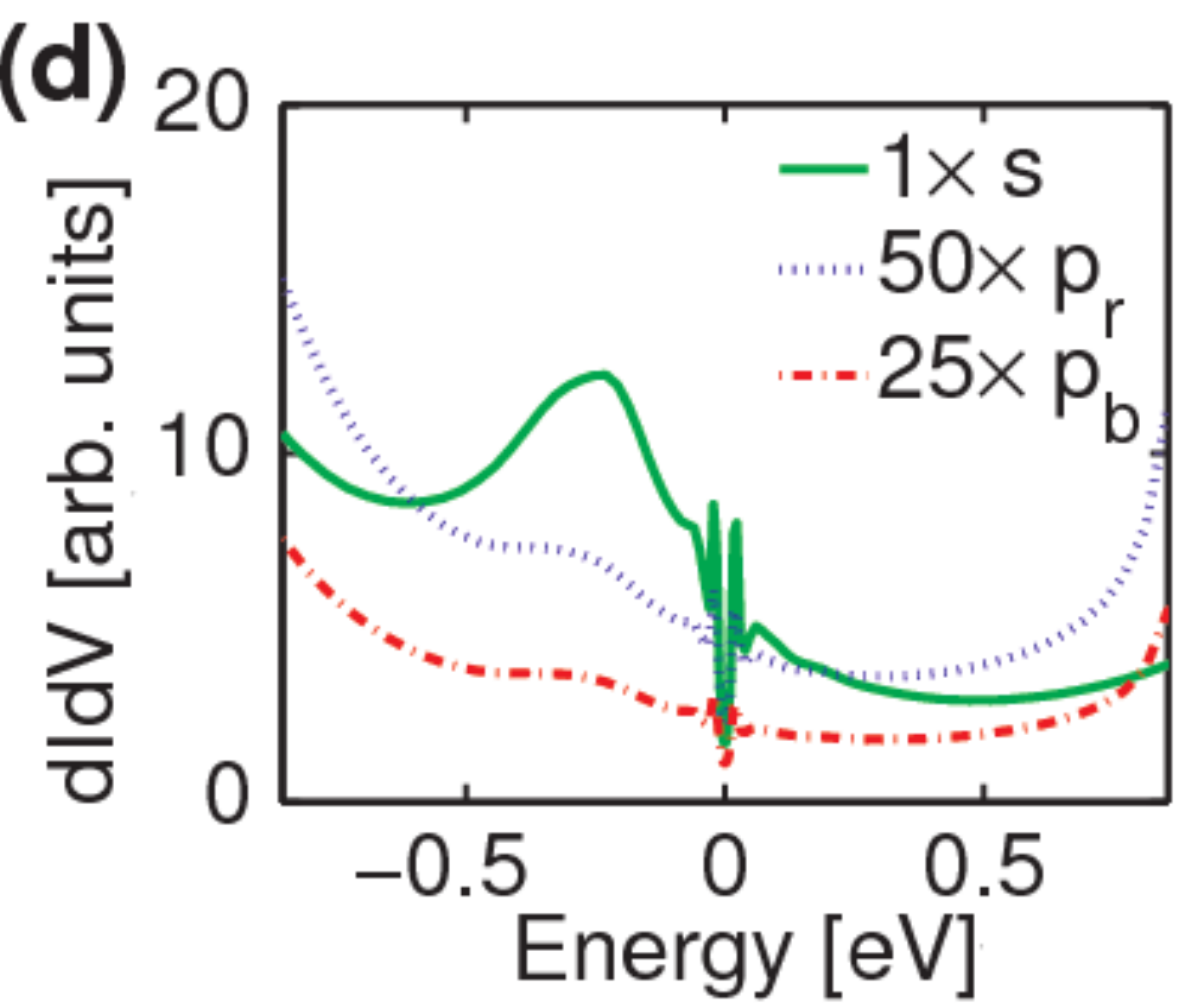}
  \end{minipage}
  \caption{(color online) (a) LDOS of $p_{z}$ of Bi, $p_{x/y}$ of
    O(Bi) and $p_{z}$ of O(Bi). The first orbital hybridizes with the
    CuO$_2$ bands, the second is dominant in BiO bands (Bi pocket
    states), and the third is rather localized; (b), (c) and (d) show
    the dI/dV curves with different tip orbitals at the high symmetry
    sites B, H and T, respectively.}
  \label{LDOS}
\end{figure}

The shape and intensity of a $dI/dV$ spectrum depend on the tip
symmetry as well as its position. In Figs. \ref{LDOS}(b-d), the
$dI/dV$ curves are calculated at $5\AA$ vertical distance between the
tip and the surface atoms. At this tip-surface distance, the gap
features are clearly observed, and calculations indicate that the
signal predominantly comes through the $p_{z}$ orbitals of the Bi
atoms.  There are clear changes in the VHS features at $\sim$-0.15
to~-0.4~eV with tip site and orbital symmetry: they seem to vanish for
the s-wave tip at a hollow site, and for both the horizontal p-wave
tip orbitals at a top site. The apparent reason for this is the
increasing spectral weight of the BiO bands with increasing binding
energy. These configurations seem to favour a relatively large overlap
between the tip orbital and the p-orbitals of O(Bi) atoms. For an
$s$-wave tip on the top site, the $dI/dV$ curve resembles the Bi $p_z$
LDOS, as expected.  However, this is also true for the horizontal
$p$-wave tips at the bridge or hollow site.  These results support the
conventional assumption that the experimental tips are predominantly
$s$-wave.

We note also a trend in the overall intensities of the spectra. As
expected, horizontal p-symmetry of the tip orbital tends to decrease
spectral intensity due to decreasing overlap.  For s-wave symmetry,
the intensity is highest for the top site and the lowest for the
hollow site. For the horizontal p-symmetry, the top site gives the
lowest intensity due to antisymmetry between $p_{z}$ of Bi and the
tip-orbital. Moving the tip either to bridge or hollow site allows an
increased overlap between the tip orbital and $p_{z}$ of Bi.

\subsection{Impurity mapping and subsurface irregularities}

dI/dV spectra are also used to detect and analyze hidden subsurface
irregularities. These can include substitutional impurities, such as
Zn or Ni replacing a Cu atom \cite{Hudson}, or dopant atoms like
oxygen, the fingerprints of which have been discussed both
experimentally \cite{McElroy} and theoretically \cite{Hirschfeld}.
However, perhaps the most fascinating potential application may be the
$C_{4} \rightarrow C_{2}$ nematic intracell symmetry breaking of the
subsurface electronic structure observed at very low values of hole
doping in the pseudogap phase\cite{Kohsaka2007}. This phase transition
in the electronic order is not observed in the topographic images, but
rather in $dI/dV$ maps and in the so called Z-maps derived from the
differential conductance data.  The relevance to the present
theoretical considerations arises from the fact that the subsurface
bonding oxygen atoms in CuO$_{2}$ appear non-equivalent when this
low-symmetry phase is mapped. We propose that breaking the symmetry of
the tip by functionalization would give additional information about
the nematic phase. Note that the observed patterns are difficult to
interpret due to the filtering effect of the BiO and SrO layers.  It
is interesting therefore to explore whether different tip symmetries
can provide insight by probing different tunneling channels. In
earlier work\cite{nieminenPRB} we showed that the bonding oxygens of
the CuO$_{2}$ layers are invisible to a spherically symmetric tip
placed right above that atom.  Hence we choose a local perturbation at
this bonding oxygen.  The simplest kind of perturbation is a change of
the onsite energy, which could be induced by a be nearby dopant atom.

Figure \ref{channels}(d) shows a tunneling tip positioned above a
bonding oxygen in the cuprate layer, equivalent to the bridge site of
Fig.~1.  To simulate the effect of a nearby impurity, we modify the
onsite energy of the $p$-orbital along the bond by -2.0
eV.\cite{perturfoot}
 In Fig. \ref{channels}(c) we compare its LDOS
(blue line) to that of the regular site (red line). The dI/dV spectra
from the bridge site with and without onsite modification are plotted for
$s$ (Fig. \ref{channels} a) and $p_{b}$ (\ref{channels}(b)) type tips
to demonstrate how the impurity effect modifies the spectrum in
different tunneling channels.  Although the difference between the two
LDOS results is striking (Fig. \ref{channels}(c)), the spectra
obtained by a rotationally symmetric tip do not reflect this
difference. This can be explained by the results of
Ref. \onlinecite{nieminenPRB}, where we showed that two tunneling
channels in antiphase are open between the oxygen and the tip
(Fig. \ref{channels}(d)), causing destructive interference between the
corresponding signals.  [The observed signal comes from different
channels involving the Cu atoms.]

In Fig. \ref{channels}(d) we show the dI/dV curve measured by a tip
with $p_{b}$ orbital, and a clear difference is seen between the
spectra of the regular and perturbed oxygens.  The shapes of LDOS (b)
and the spectra (d) are not identical, but the relative change of the
spectrum follows that of LDOS, especially over the range between
$-0.5 - 0.0eV.$  The change in the spectrum of Fig. \ref{channels}(d) is 
caused by a reversal of the relative phase between the two tunneling
channels in Fig. \ref{channels}(a) due to p-type symmetry
of the tip, a point to which we will return below. 
This result suggests that experiments probing the breakdown of 
$C_{4}$ symmetry, such as those in Ref. \onlinecite{Kohsaka2007},
could be further clarified by varying the tip symmetry.

\section{Analysis}
\label{analysis}

\subsection{Tip symmetry and contrast changes in the STM map}

A proper understanding of the observed topographical images and
spectra requires an analysis of the filtering matrix elements $M_{tc}$
which describe the coupling between the tip and the surface
orbitals, propagation across the oxide layers, and coupling between
the superconducting cuprate layer and the covering oxide layers.
Starting from the latter, the cuprate layer is coupled to
the oxide layers mainly through the overlap between the $d_{z^{2}}$ orbitals
of the Cu atoms and the $p_{z}$ orbitals of the apical oxygen
above. The propagation through the oxide layers is mediated by
$p_{z}$ orbitals of the apical oxygen and the Bi atom above on the
surface. When the tip symmetry is varied, the relevant factor is that 
$p_{z}$ of the
surface Bi is the dominant surface orbital mediating the signal from
the cuprate layer to the microscope tip. Hence, the analysis of the
observed results can be built on the overlap between the tip orbital
and the $p_{z}$ orbitals of the Bi atoms in the vicinity of the tip.
While other surface orbitals are present, they are not strongly coupled to the 
cuprate bands, and hence they have a relatively low spectral weight in the 
vicinity of the Fermi energy. Thus, the horizontal p-orbitals of Bi
and O(Bi) atoms are the dominant orbitals of the BiO bands related to
the ``bismuth pocket'' states above Fermi-level and the corresponding
valence bands below the ``spaghetti'' of bands 
(Fig. \ref{LDOS}(a))\cite{bifoot}.

Since only the Bi $p_z$ surface orbital is important, the analysis of
the various tip orbitals is rather straightforward.  The $s-$ and
$p_{z}$-orbitals couple best on the top site, as seen in
Fig. \ref{LDOS}(b). For the horizontal tip orbitals, the
overlap between the tip and $p_{z}$ of a surface Bi is lowest when the
tip is directly above the Bi atom. In the case of $p_{r}$ both the
bridge and hollow sites seem to give a large overlap between tip
orbital and $p_{z}$ of Bi, but the hollow sites appear brighter since
the overlap between the horizontal p-orbitals contributing to BiO
bands is larger at this site.
For $p_{b}$ the fourfold symmetry is broken. Since the orbital is
oriented along the bonds between Cu atoms, there is a zero overlap
line when following the bonding direction perpendicular to the orbital
orientation. On the other hand, $p_{b}$ overlaps strongly with the two Bi
atoms at the bridge site in the antinodal direction of $p_{b}$. Moving
towards the vicinal hollow sites preserves the strong overlap. At the
hollow site, a strong overlap with four Bi atoms can be found. Since
the distance is $\sqrt{2}$-fold as compared to the bridge site, overall the
overlap remains roughly the same when moving from the bridge site to
the hollow site, resulting in bright and dark stripes in
Fig. \ref{topo}(e).

\subsection {Hidden irregularities and the relative phase between two
  tunneling channels.}

\begin{figure}[th]
  \centering
 
  \begin{minipage}{4.3cm}%
    \includegraphics[width=1.0\textwidth]{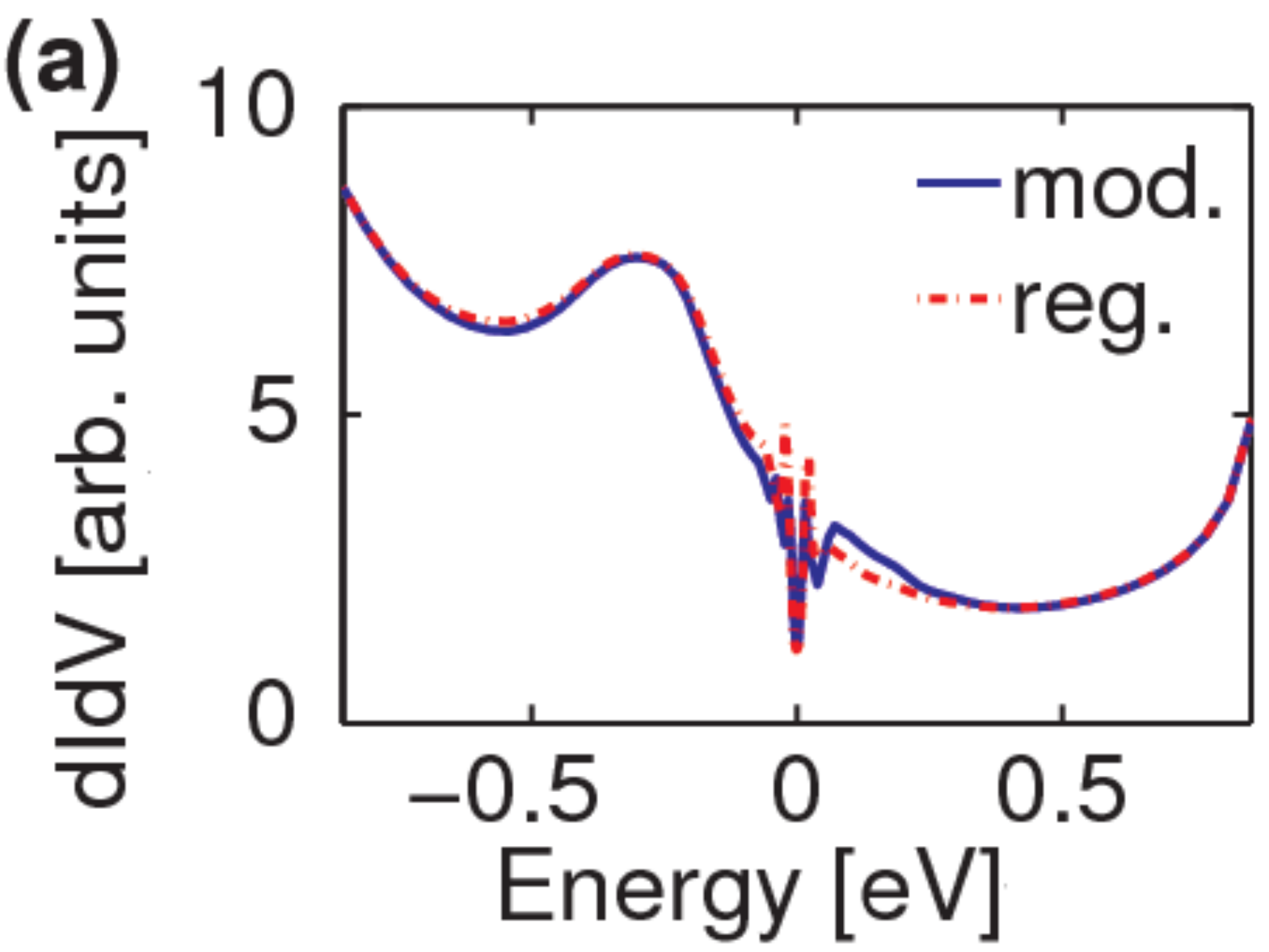}
  \end{minipage}%
  \begin{minipage}{4.3cm}%
    \includegraphics[width=1.0\textwidth]{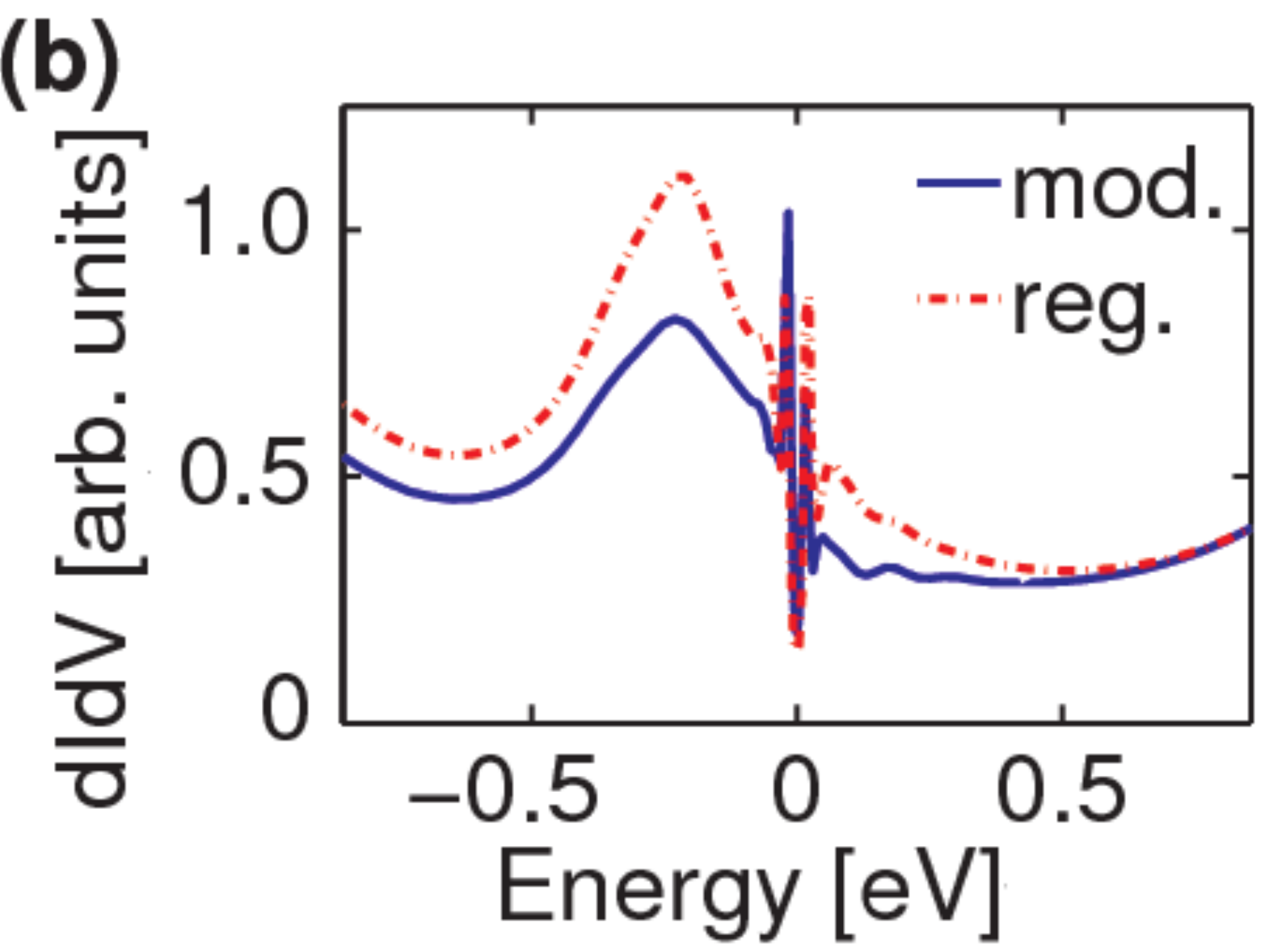}
  \end{minipage}\\
  \begin{minipage}{4.3cm}%
    \includegraphics[width=1.0\textwidth]{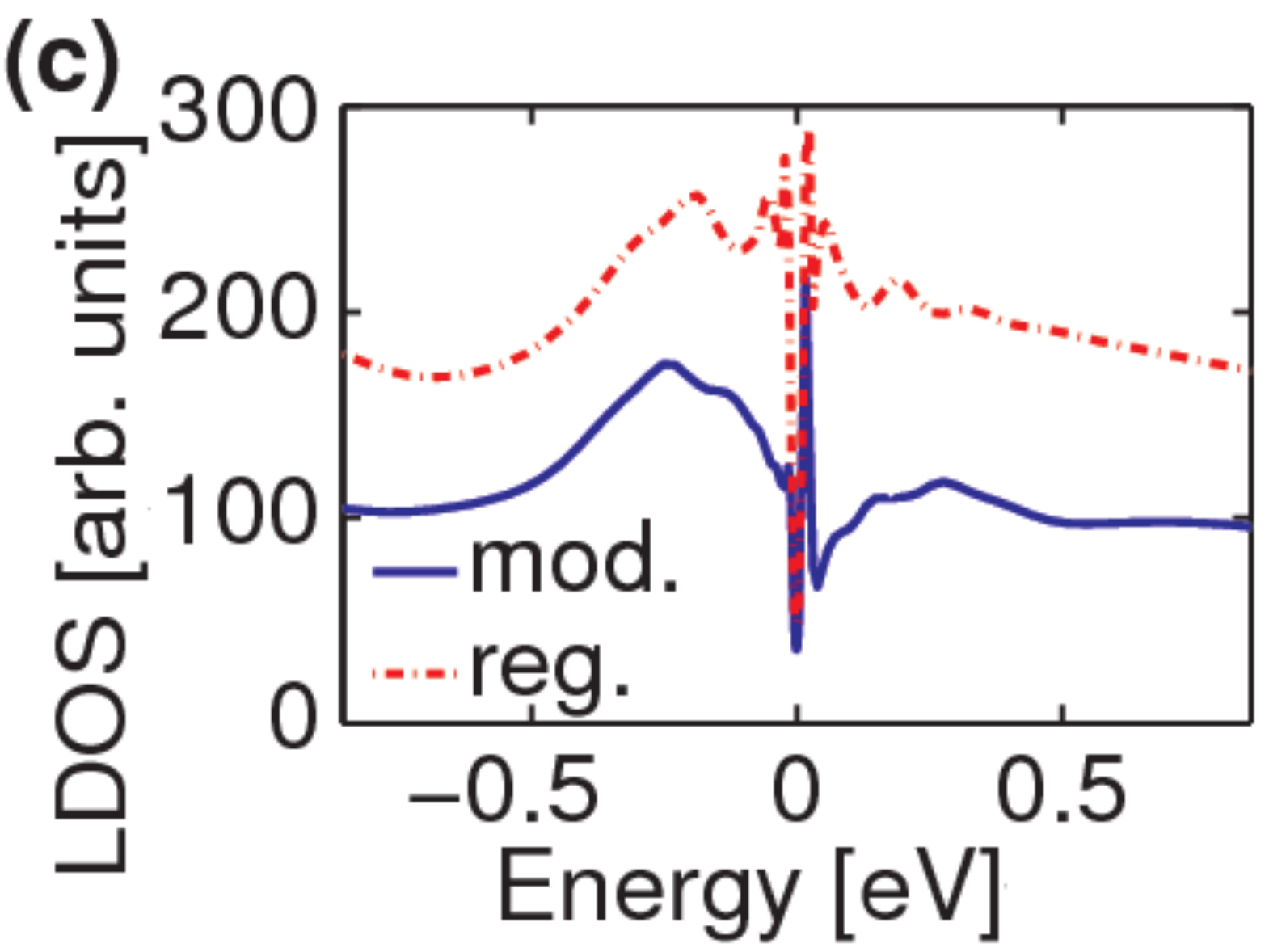}
  \end{minipage}
  \begin{minipage}{4.0cm}%
    \includegraphics[width=1.0\textwidth]{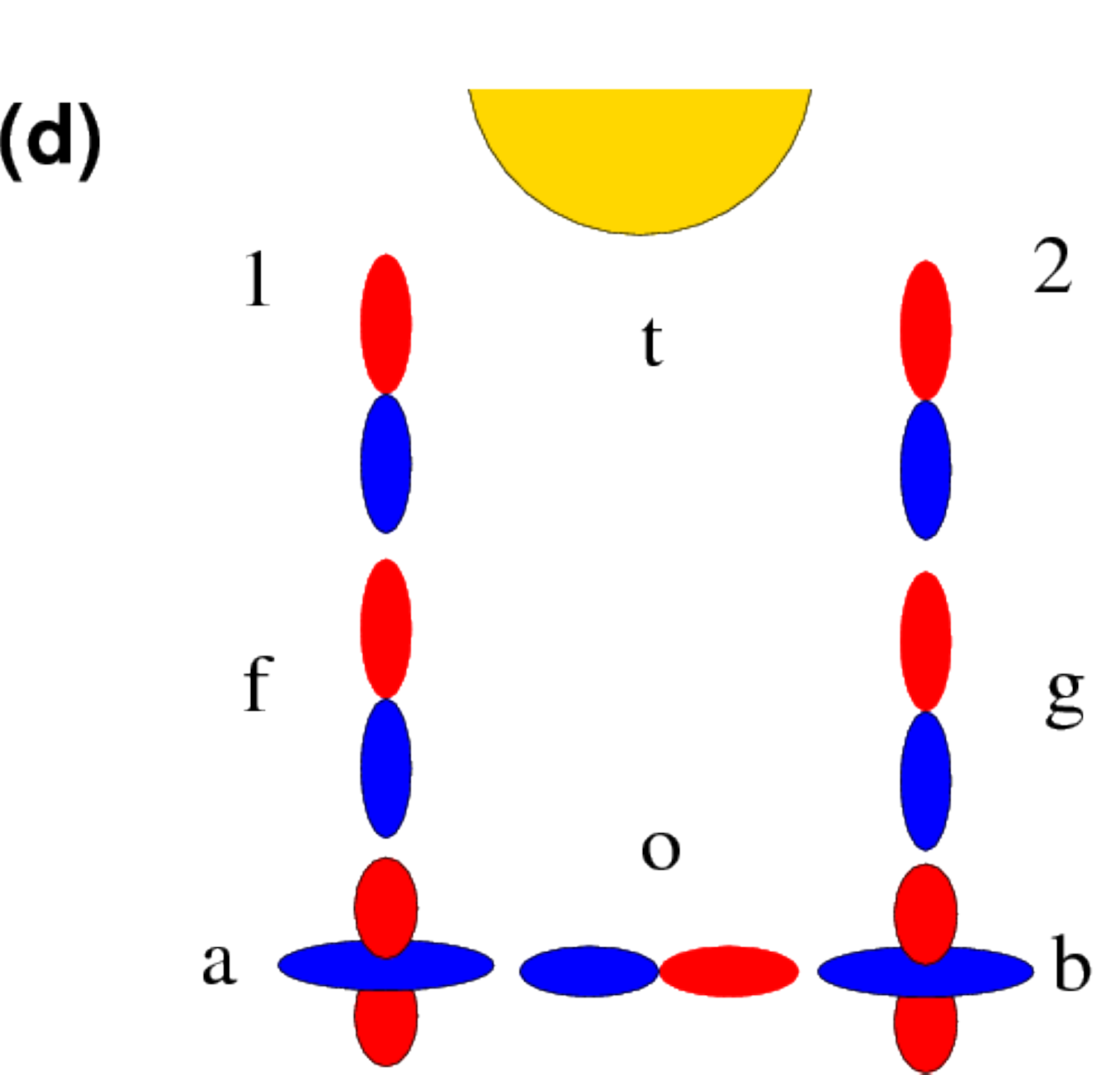}
  \end{minipage}%
  \caption{(color online) 
(a) dI/dV spectra of the regular and modified
    bonding oxygen using s-symmetric tip above the oxygen; (b) The
 same spectra using
    $p_b$-symmetric tip above the oxygen; (c) LDOS of the $p$-orbital of the
    bonding oxygen in the cuprate layer. \emph{Regular} case (red)
    is compared with an orbital with a \emph{modified} onsite energy (blue); 
    (d) Two competing channels from the
    bonding oxygen ($o$) through $d_{z}$ orbitals ($a$ or $b$),
    apical oxygens ($f$ and $g$) and the surface bismuths ($1$ and
    $2$) to the tip ($t$).}
  \label{channels}
\end{figure}

We consider the possibility of analyzing subsurface irregularities by
functionalizing the tip. This may be possible if the interference
between two different tunneling channels can be controlled.  As an 
example, let
us consider the situation where the tip is at the bridge site (see
Fig. \ref{channels} (d)), i.e., above an oxygen atom in the CuO$_2$
layer at an equal distance from the two neighbouring
Bi atoms ($1$ and $2$). This opens two channels from the CuO$_2$ layer
($d_{z2}$ orbitals of Cu atoms $a$ and $b$) with filtering matrix
elements $M_{1a}=V_{t1}G^{0}_{1f}V_{fa}$ and
$M_{2b}=V_{t2}G^{0}_{2g}V_{gb}$ which are equal apart from a phase factor.  
It
is obvious that $M_{1a}$ and $M_{2b}$ are equal in the case of $s$- or
$p_{z}$-symmetric tip, since in that case $V_{t1} = V_{t2}$, but the
opposite sign is obtained with horizontal $p$-orbitals.

To consider interference effects, we also need to compare elements of
the spectral function 
\begin{eqnarray*}
  S_{aoa} & = & G_{ao}^{+}\Sigma''_{oo} G_{oa}^{-} \\ 
  S_{bob} & = & G_{bo}^{+}\Sigma''_{oo} G_{ob}^{-} \\ 
  S_{aob} & = & G_{ao}^{+}\Sigma''_{oo} G_{ob}^{-}, 
\end{eqnarray*}
%
%
where $o$ denotes the
p-orbital of the oxygen along the bond between the Cu atoms. It is
evident that the p-symmetry leads to an antiphase between elements
$G_{ao}$ and $G_{ob}$, and hence the two first spectral terms are
equal to each other, but the last one has the opposite sign.

Finally, using equations \eqref{mconductance}, \eqref{filter} and
\eqref{dmatrix} we obtain the total current from orbital $o$ to the
tip:

\begin{equation}
\begin{split}
  I_{o} = & 
\rho_{tt}(\vert M_{1a} \vert^{2} S_{aoa} + \vert M_{2b} \vert^{2} S_{bob} + \\
& M_{1a}S_{aob}M^{\dagger}_{b2} + M_{2b}S_{boa}M^{\dagger}_{a1}) \\
 = &
\rho_{tt}(2 \vert M_{1a} \vert^{2}  \mp 2\vert M_{1a} \vert^{2})S_{aoa}.
\end{split}
\end{equation}


It is clear that '-' sign on the right hand side of Eq. (6) applies to
the case with $V_{t1} = V_{t2}$, i.e., the $s$- or
$p_{z}$-symmetry. Hence the oxygen below the tip would be invisible in
STM. On the other hand, constructive interference is obtained for
horizontally $p$-symmetric tip. This should emphasize any variations
in the horizontal p-orbital of the bonding oxygen. The situation is of
course reversed if the bonding oxygen is replaced by an $s$-wave
symmetric perturbation, i.e. there is a constructive interference for
$s$- and $p_{z}$-symmetric tips.  This explains our comparison of a
spectrum at the bridge site for unperturbed system with a vacancy at
the site of the bonding oxygen.

With regard to the $C_{4} \rightarrow C_{2}$ transition, the main
practical conclusion of our analysis is that obtaining information
related to the non-equivalent bonding oxygens can be enhanced by
manipulating the tip symmetry by functionalizing molecules.  This
suggests that an s-wave symmetric tip gathers the signal from
a neighbourhood of the bonding oxygen below, while a
p-wave tip should be more selective to the on-site oxygen orbitals.

\section{Conclusions}

We have investigated the effect of different tip symmetries on
topographic STM images and dI/dV spectra.  The change in tip symmetry
can lead to contrast inversion and even to the breaking up of the $C_{4}$
symmetry due to strong differences in the nature of the tip-surface
overlap. Furthermore, the corrugation curves show that the horizontal
p-orbitals at the apex of the tip would be more sensitive to position
of the surface atoms. The dI/dV curves reveal the leading sample 
states (CuO$_2$-band,
BiO-band or some more localized states) to which different symmetries tend
to couple. More strikingly, changing the symmetry of the tip can be used to 
change the relative phase between two open tunneling channels. This
could be applied to detect subsurface irregularities which would otherwise
be invisible to STM.

{\bf Acknowledgments} This work is supported by the US Department of
Energy grant DE-FG02-07ER46352 and benefited from the allocation of
supercomputer time at NERSC and Northeastern University's Advanced
Scientific Computation Center (ASCC).  I.S would like to thank Ulla
Tuominen Foundation for financial support.  This work benefited from
resources of Institute of Advanced Computing, Tampere. Discussions
with Matti Lindroos are gratefully acknowledged.


\begin{thebibliography}{99}
\bibitem{abfoot1} Matrix element effects are well-known to be
  important in other highly resolved spectroscopies such as
  angle-resolved photoemission\cite{newarpes}, resonant inelastic
  x-ray scattering(RIXS)\cite{rixs}, Compton scattering \cite{Compton}
  and positron-annihilation \cite{positron}.
\bibitem{selffoot} In the present calculations, we apply a self-energy
  adapted from ref. \onlinecite{tanmoyop} to the $d_{x^{2}-y^{2}}$
  orbitals of Cu atoms in order to obtain a realistic dispersion
  renormalization.
\bibitem{abfoot2} We expect the rigid band model to be a good
  approximation for doping away from the CuO$_2$ planes. It will be
  interesting to further examine doping effects by using supercell or
  other first principles approaches \cite{new1,new2}.
\bibitem{tunnelingfoot} This equation can also be derived as a special
  case of Landauer-B\"uttiker approach as expressed in
  Ref.\onlinecite{meirwingreen}.  On the other hand,
  Eq. \eqref{conductance} can be converted into
  Tersoff-Hamann\cite{TH} formalism in the limiting of a spherical tip
  orbital in real space\cite{korventausta}.
\bibitem{hoppingfoot} Note that the density matrices in
  Eq. \eqref{conductance} concern, in principle, an interacting system. An
  alternative approach would be to use noninteracting density matrices and
  replace the hopping terms by a t-matrix, $T=V(1-GV)^{-1}.$ However,
  for the tip distance used in this work, the tip-sample hopping terms are
  of the order of $0.01-0.02eV$, while the interaction terms within
  the sample are of the order of $1eV$ [e.g., the hopping
  between $p_{b}$ of the bonding oxygen and the $d_{x^{2}-y^{2}}$ of
  the neighbouring Cu atom is $1.62 eV].$ Hence, it is safe to utilize
  noninteracting density matrices in calculating the conductance.
\bibitem{ldafoot} LDA calculations tend to give conducting BiO bands
  the bottom of which are slightly below Fermi-level. However, one
  never observes these states in ARPES or STS, and hence it is
  possible that doping the Bi2212 samples may lift these states off
  the Fermi level \cite{Hsin}.
\bibitem{perturfoot}We tested several values for the
onsite energy, but this value was chosen to best exemplify
the effect of a local perturbation at the bonding oxygen site.
\bibitem{bifoot} $p_{z}$ of O(Bi) should be rather weakly hybridized
  with any orbitals; possibly it is very localized and hence gives
  only marginal contributions to the spectra.
\end{thebibliography}
\end{document}